\documentclass[%
superscriptaddress,
 amsmath,amssymb,
 aps,
jcp
]{revtex4-2}

\usepackage{graphicx}
\usepackage{dcolumn}
\usepackage{bm}
\usepackage{color}
\usepackage[dvipsnames]{xcolor}
\usepackage[colorlinks,citecolor=blue]{hyperref}

\newcommand{\gcc}{g$\,$cm$^{-3}$}
\definecolor{darkviolet}{rgb}{0.58, 0.0, 0.83}

\begin{document}

\preprint{APS/123-QED}

\title{Nonideal Mixing Effects in Warm Dense Matter \\ Studied with First-Principles Computer Simulations }

\author{Burkhard Militzer}
\email{militzer@berkeley.edu}
\affiliation{Department of Earth and Planetary Science, University of California, Berkeley, CA 94720, USA}
 \affiliation{Department of Astronomy, University of California, Berkeley, CA 94720, USA}

\author{Felipe Gonz\'alez-Cataldo}
\affiliation{Department of Earth and Planetary Science, University of California, Berkeley, CA 94720, USA}

\author{Shuai Zhang}
\affiliation{Laboratory for Laser Energetics, University of Rochester, Rochester, NY 14623, USA}

\author{Heather D. Whitley}
\affiliation{Lawrence Livermore National Laboratory, Livermore, California 94550, USA}

\author{Damian C. Swift}
\affiliation{Lawrence Livermore National Laboratory, Livermore, California 94550, USA}

\author{Marius Millot}
\affiliation{Lawrence Livermore National Laboratory, Livermore, California 94550, USA}

\begin{abstract}
We study nonideal mixing effects in the regime of warm dense matter (WDM) by computing the shock Hugoniot curves of BN, MgO, and MgSiO$_3$. First, we derive these curves from the equations of state (EOS) of the fully interacting systems, which were obtained using a combination of path integral Monte Carlo calculations at high temperature and density functional molecular dynamics simulations at lower temperatures. We then use the ideal mixing approximation at constant pressure and temperature to rederive these Hugoniot curves from the EOS tables of the individual elements. We find that the linear mixing approximation works remarkably well at temperatures above $\sim$$2 \times 10^5\,$K, where the shock compression ratio exceeds $\sim$3.2. The shape of the Hugoniot curve of each compound is well reproduced. Regions of increased shock compression, that emerge because of the ionization of L and K shell electrons, are well represented and the maximum compression ratio on the Hugoniot curves is reproduced with high precision. Some deviations are seen near the onset of the L shell ionization regime, where ionization equilibrium in the fully interacting system cannot be well reproduced by the ideal mixing approximation. This approximation also breaks down at lower temperatures, where chemical bonds play an increasingly import role. However, the results imply that equilibrium properties of binary and ternary mixtures in the regime of WDM can be derived from the EOS tables of the individual elements. This significantly simplifies the characterization of binary and ternary mixtures in the WDM and plasma phases, which otherwise requires large numbers of more computationally expensive first-principles computer simulations.
\end{abstract}

\keywords{Linear mixing, equation of state, Warm Dense Matter, shock Hugoniot curves.}

\maketitle
\section{Introduction}

The physical properties of hot, dense plasmas have been studied with experimental and theoretical techniques for decades~\cite{Ebeling1991} because their behavior is important for a number of energy technologies, including inertial confinement fusion (ICF)~\cite{Gaffney2018,Betti2016,Miyanishi2015,Lindl2014,Hammel2010,Seidl2009}. On the path to fusion, the sample material typically passes through the regime of warm dense matter (WDM), which encompasses matter at solid-state densities and elevated temperatures of 10$^4$–10$^7$ K. This regime is particularly difficult to describe with theoretical methods because the densities are too high and interaction effects are too strong for typical plasma theory to be applicable~\cite{Ro86,Ro90} or for Saha ionization models to work properly~\cite{Ebeling1976}. On the other hand, the temperatures are too high and the fraction of excited electrons too large for conventional condensed matter theory to be applicable. The temperature is also not high enough for screening effects to become the dominant type of interaction and thus Debye plasma models~\cite{Debye1923} do not work well. All particles are strongly interacting, which renders the system nonideal. There is no small parameter that would allow for analytical descriptions to be appropriate. Chemical bonds still play a role, even though they are typically short-lived. The electrons may be highly excited and partially ionized. Pauli exclusion effects are relevant when the ionization equilibrium is established, which renders the system partially degenerate. A good fraction of the electrons occupy core states because density is orders of magnitude too low for them to form a rigid neutralizing background. In this regard, a one-component plasma model would be a poor description of WDM. Despite these challenges, the development of a rigorous and consistent theoretical framework to describe WDM remains of high importance. 

Significant progress has been made with laboratory experiments and first-principles computer simulations. A number of different simulation methods have been advanced~\cite{Graziani2014Book}. These simulation methods enable us to compute the equation of state (EOS) of materials over a wide range of conditions that are also relevant in astrophysics. In the interiors of giant planets, for example, not only hydrogen-helium mixtures but also rocky materials are exposed to tens of megabars and $\sim$10$^4$ K~\cite{Militzer2019b,Wahl2017,Soubiran2017,Militzer2016b,Soubiran2016,Gonzalez-Cataldo2014,Gonzalez-Cataldo2016,Wahl2013a,Wilson2012,Militzer2013c}. Accurate EOSs are needed to characterize their interior structure and evolution~\cite{Militzer2016b,Baraffe2014}. The discovery of thousands of exoplanets with ground-based observations and space missions as well as the unexpected diversity in their masses and radii~\cite{Guillot1999,Deming2020} considerably broadened the range of conditions and materials of interest~\cite{Madhusudhan2011,Wagner2012,Wilson2014}. 

In the interior of stars, matter is exposed to a wide range of temperatures $\sim$10$^4-$10$^8\,$K. The most detailed constraints on the interior conditions come from measurements of normal mode oscillations of our Sun~\cite{Christensen1996,Christensen2002,Schumacher2020}. Such astero-seismological observations are also employed to constrain the interiors of distant stars~\cite{Aerts2019}. 
The interpretation of these observations would not be possible without a comparable development of laboratory experiments to probe such extreme conditions~\cite{Fortney2009b}, which employ a variety of high-velocity impacts~\cite{Nellis1991,Weir1996}, lasers~\cite{Craxton2015,Pape2018,Remington2015,Yabuuchi2019,Zhang2018,Betti2016,Seidl2009,Miyanishi2015}, and magnetic compression techniques~\cite{Knudson2012,Gomez2020}. The goal of this article is to support these activities by providing theoretical methods to predict the state of WDM with computer simulations and to derive the EOS of materials over a wide range of temperature-pressure conditions. This will aid the interpretation of current experiments or help with their design in the future. Therefore any material and condition, that can be probed with current experimental facilities, is of potential interest. 

The range of pressure-temperature conditions of interest is very large and so is the space of possible chemical compositions. The challenge of dealing with this huge number of materials and conditions is not unique to the field of WDM~\cite{materials_project}. For binary and ternary mixtures, the number of relevant conditions scales as $N_\rho \times N_T \times N_{\rm E}^2 \times N_{\rm mix}$ and $N_\rho \times N_T \times N_{\rm E}^3 \times N_{\rm mix}^2$ where $N_\rho, N_T, N_{\rm E},$ and $N_{\rm mix}$, respectively, are the numbers of densities, temperatures, elements, and mixing ratios of interest. The resulting numbers are often so large that an exhaustive coverage is impractical not only for laboratory experiments but also for computer simulations, although exceptions exist~\cite{ZhangCH2018}. To simplify the computation of WDM, we investigate the validity of the linear mixing approximation at high pressures and temperatures in this article.

The linear mixing approximation is a widely used approach to obtain the equation of state (EOS) of materials from the individual components if the information about the fully interacting mixture is lacking. The ideal mixing rule is often used to study gaseous mixtures of simple elements, such hydrogen, helium, carbon and oxygen, to understand the atmospheres and interior of gas giant planets, where complex mixtures at different concentrations emerge and whose physical properties are unknown~\cite{Chabrier1990,Ross1998,Bethkenhagen2017}. The accuracy of this approach depends largely on the thermodynamic conditions at which it is applied~\cite{Mi05}. A wide variety of systems has been explored under the assumption of ideal mixing~\cite{Li2018,Saumon2004,Deng2019,Baraffe2008,Guillot1999a,Morales2009,Militzer2009,French2012}. The hypothesis works remarkably well for hydrocarbon mixtures in the warm dense matter regime~\cite{ZhangCH2017}, water-hydrogen mixtures~\cite{Soubiran2015}, as well as in hydrogen-helium mixtures enriched in heavier elements~\cite{Hubbard2016,Militzer2013b,Soubiran2016,Fortney2009b}.
The range of validity of the linear mixing hypothesis has been explored in great detail for hydrogen-helium mixtures~\cite{McMahon2012}. As an example of a nonlinear mixing effect, Vorberger \emph{et al}.~\cite{Vorberger2007} showed that the presence of helium stabilizes the hydrogen molecules in the mixture and thus moves their molecular-to-metallic transition to higher pressures.  

In this article, we employ two first-principles computer simulation methods, path integral Monte Carlo (PIMC) calculations and density functional molecular dynamics (DFT-MD) simulations, to study nonideal mixing effects in the regime of WDM. With this approach, we have been able to produce several EOSs in previous years, which have supported the laboratory experiments. In particular, our simulation results for warm dense carbon~\cite{Benedict2014} are currently being used to design NIF targets. 

In this article, we will show with our first-principles computer simulations that the compositional dependence of the EOS is manageable in the regime of WDM. We will demonstrate that the shock Hugoniot curves of various mixtures can be derived with good accuracy for temperatures above $\sim 2 \times 10^5\,$K by invoking the ideal mixing approximation at constant pressure and temperature. This means chemical bonds between species no longer play an important role at these temperatures. It is still surprising that the properties of hot, dense MgSiO$_3$ can be derived from those of the pure substances because one essentially assumes, e.g., a Mg ion in a dense MgSiO$_3$ environment behaves very similarly as one that is surrounded by other Mg ions at the same $P$-$T$ conditions. Without verification, there is no guarantee that the degree of ionization will be similar in the two systems. The goal of this article is to investigate these similarities and to characterize the nonlinear mixing effects quantitatively for the three representative WDM materials BN, MgO, and MgSiO$_3$.

\section{Methods and Assumptions}

\begin{figure}
    \centering
    \includegraphics[width=0.65\columnwidth]{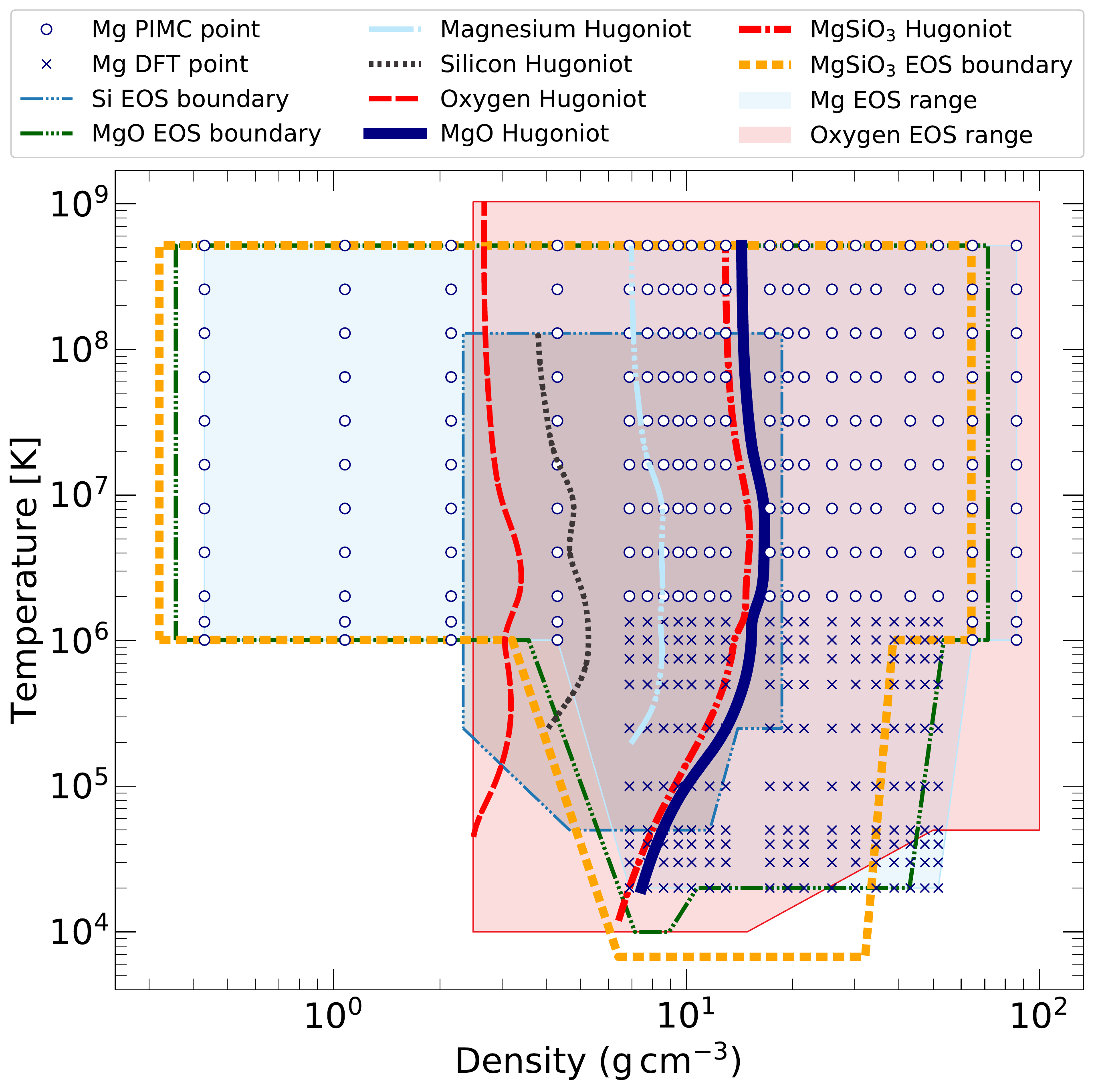}
    \includegraphics[width=0.65\columnwidth]{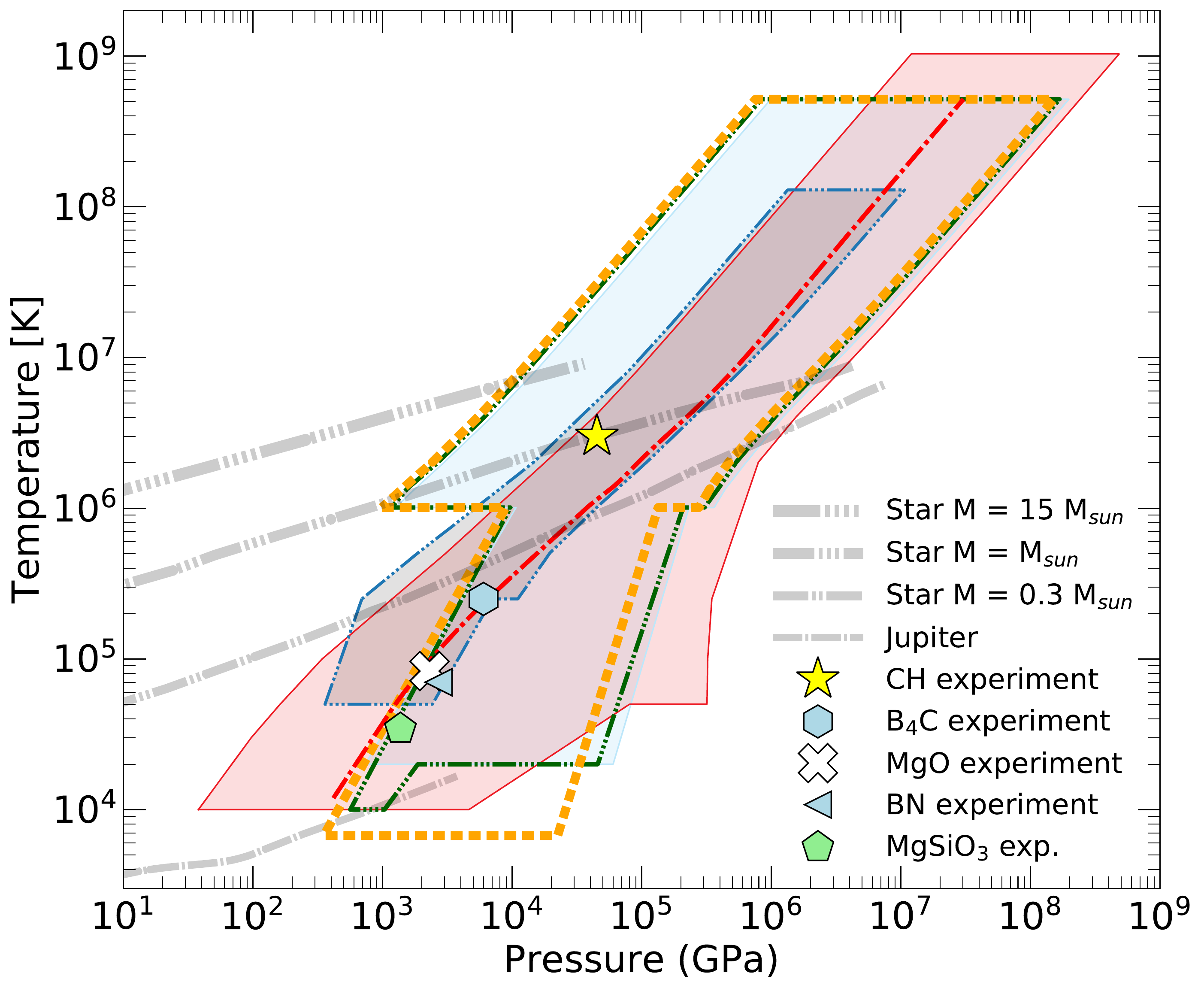}
    \caption{The boundaries of our EOS tables are shown in density-temperature and in pressure-temperature spaces. Various shock Hugoniot curves are included and in the case of magnesium, we also show the individual EOS points that were computed with PIMC simulations (circles) for T$\ge 2 \times 10^6\,$K and with DFT-MD (crosses) for lower temperatures. In the lower panel, interiors conditions of Jupiter~\cite{Wahl2017a} and stars of different masses~\cite{SC95} have been added but some lines and symbols from the upper panel have been omitted for clarity while we marked the maximum pressures in recent experiments with CH~\cite{Kritcher2020}, B$_4$C~\cite{Zhang2020}, MgO~\cite{McCoy2019}, BN~\cite{ZhangBN2019}, and MgSiO$_3$~\cite{Millot2020}. The corresponding temperatures were not measured but derived with simulations.}
\label{fig:EOS_ranges}
\end{figure}

We derived the equation of state of every material under consideration, Mg, Si, O, B, N, MgO, MgSiO$_3$, and BN, by performing series of first-principles computer simulations that employed PIMC simulations at high temperature and standard Kohn-Sham DFT-MD calculations at low temperature. We describe these methods in the following two sections. 

\subsection{PIMC simulations}

Path integral Monte Carlo (PIMC) methods have gained considerable interest as a state-of-the-art, stochastic first-principles technique to compute the properties of interacting quantum systems at finite temperature. This formalism results in a highly parallel implementation and an accurate description of the properties of materials at high temperature where the electrons are excited to a significant degree~\cite{Mi06,Benedict2014,Driver2015,Hu2016,ZhangBN2019}. 
The application of the PIMC method to light elements from hydrogen through neon~\cite{Driver2015} has been possible due to the development of free-particle nodes~\cite{Ce95,Ce96} while simulations of heavier elements relied on the advancement of Hartree-Fock nodes~\cite{MilitzerDriver2015}. The latter approach enables one to efficiently incorporate localized electronic states into the nodal structure, which extends the applicability of the path integral formalism to heavier elements and lower temperatures~\cite{ZhangSodium2017,Driver2018}. Furthermore, PIMC treats all electrons explicitly and avoids the use of pseudopotentials. The PIMC simulation time scales as $1/T$, proportional to the length of the paths, which is efficient at high-temperature conditions, where most electrons including the K shell are excited. Predictions from PIMC simulations at intermediate temperatures have been shown to be in good agreement with predictions from density functional theory molecular dynamics (DFT-MD) simulations~\cite{Mi09,ZhangCH2018}.

The fundamental techniques for the PIMC simulations of bosonic systems
were developed in Ref.~\cite{PC84} and reviewed in Ref.~\cite{Ce95}.
Subsequently the algorithm was extended to fermionic systems by introducing the {\em restricted} paths approach~\cite{Ce91,Ce92,Ce96}.
The first results of this simulation method were
reported in the seminal work on liquid $^3$He~\cite{Ce92} and dense
hydrogen~\cite{PC94}. In subsequent articles, this method was applied to
study hydrogen~\cite{Ma96,Mi99,MilitzerThesis,MC00,MC01,Mi01},
helium~\cite{Mi06,Mi09,Mi09b}, hydrogen-helium mixtures~\cite{Mi05}
and one-component plasmas~\cite{JC96,MP04,MP05}. In recent years, the
PIMC method was extended to simulate plasmas of various first-row
elements~\cite{Benedict2014,DriverNitrogen2016,Driver2017,ZhangCH2017,ZhangCH2018,Zhang2018}
and with the development of Hartree-Fock nodes, the simulations with heavier nuclei up to silicon became
possible~\cite{MilitzerDriver2015,Hu2016,ZhangSodium2017,Driver2018}.

The PIMC method is based on the thermal density matrix of a quantum
system, $\hat\rho=e^{-\beta \hat{\cal H}}$, that is expressed as a
product of higher-temperature matrices by means of the identity
$e^{-\beta \hat{\cal H}}=(e^{-\tau \hat{\cal H}})^M$, where
$M$ is an integer and $\tau\equiv\beta/M$ represents the time step of a path integral in
imaginary time. The path integral emerges when the operator $\hat\rho$
is evaluated in real space,
\begin{equation}
\left<\mathbf R|\hat\rho| \mathbf R'\right>=\frac{1}{N!}\sum_{\mathcal P}(-1)^{\mathcal P}\oint_{\mathbf R\to\mathcal P\mathbf R'}\mathbf{dR}_t\, e^{-S[\mathbf R_t]}.
\label{PI}
\end{equation}
The sum includes all permutations, $\mathcal P$, of
$N$ identical fermions in order project out the antisymmetric
states.  For sufficiently small time steps, $\tau$, all many-body
correlation effects vanish and the action, $S[\mathbf R_t]$, can be
computed by solving a series of two-particle
problems~\cite{PC84,Na95,BM2016}. The advantage of this approach is that 
all many-body quantum correlations are recovered
through the integration over paths. The integration also enables
one to compute quantum mechanical expectation values of thermodynamic
observables, such as the kinetic and potential energies, pressure,
pair correlation functions and the momentum
distribution~\cite{Ce95,Militzer2019}. Most practical implementations of
the path integral techniques rely on Monte Carlo sampling techniques
because the integral has $D \times N \times M$ dimensions in
addition to sum over permutations. ($D$ is the number of spatial dimensions.) The method becomes
increasingly efficient at high temperature because the length
of the paths scales like $1/T$. In the limit of low temperature, where
few electronic excitations are present, the PIMC method becomes
computationally demanding and the Monte Carlo sampling can become inefficient.
Still, the PIMC method avoids any exchange-correlation approximation
and the calculation of single-particle eigenstates, which are embedded 
in all standard Kohn-Sham DFT calculations. 

The only uncontrolled approximation within fermionic PIMC calculations
is the use of the fixed-node approximation, which restricts the paths
in order to avoid the well-known fermion sign
problem~\cite{Ce91,Ce92,Ce96}. Addressing this problem in PIMC is
crucial, as it causes large fluctuations in computed averages due to
the cancellation of positive and negative permutations in
Eq.~\eqref{PI}. We solve the sign problem approximately by restricting
the paths to stay within nodes of a trial density matrix that we obtain 
from a Slater determinant of single-particle density matrices,
\begin{equation}
\rho_T({\bf R},{\bf R'};\beta)=\left|\left| \rho^{[1]}(r_{i},r'_{j};\beta) \right|\right|_{ij}\;,
\label{FP}
\end{equation}
that combined free and bound electronic states~\cite{MilitzerDriver2015,Driver2018}, 
\begin{eqnarray}\label{rho1}
\rho^{[1]}(r,r';\beta) &=& \sum_{k} e^{-\beta E_k} \, \Psi_k(r) \, \Psi_k^*(r')\\
 &+& \sum_{I=1}^{N} \sum_{s=0}^{n} e^{-\beta E_s} \Psi_s(r-R_I) \Psi_s^*(r'-R_I)\;.
\quad.
\end{eqnarray}
The first sum includes all plane waves, $\Psi_k$ while the second represents $n$ 
bound states $\Psi_s$ with energy $E_s$ that are localized around all atoms $I$. 
Predictions from various slightly differing forms of this approach have been 
compared in Ref.~\cite{ZhangSodium2017} 

The PIMC simulations were performed with the CUPID code~\cite{MilitzerThesis} using periodic boundary conditions. For pure O, Mg, Si, and N systems, we considered simulation cells with 8 nuclei and 64, 96, 112, and 56 electrons, respectively. For the simulations of boron, slightly larger cells with 30 nuclei and 150 electrons are used. For MgO, we considered 80 electrons, 4 Mg, and 4 O nuclei. For BN, we considered 144 electrons, 12 B, and 12 N nuclei while our MgSiO$_3$ simulations consisted of 3 Mg, 3 Si, and 9 oxygen nuclei as well as 144 electrons. Finite size effect were discussed when we first reported these EOS calculations. A detailed finite-size study is provided in the supplementary material of Ref.~\cite{Driver2015} that showed that finite size effects are relatively small because the most important changes in the energy and pressure are caused by the ionization of various electronic states. While the ionization equilibrium depends on the thermodynamic conditions of the plasma, it does not require large simulation cells to capture these effects. 

We enforced the nodal constraint in small steps of imaginary time of $\tau=1/8192$ Ha$^{-1}$, while the pair density matrices~\cite{Militzer2016} were evaluated in steps of 1/1024 Ha$^{-1}$. This results in using between 2560 and 5 time slices for the temperature
range that was studied with PIMC simulations. These choices converged
the internal energy per atom to better than 1\%.  We have shown the
associated error is small for relevant systems at sufficiently high
temperatures~\cite{Driver2012}.

\subsection{DFT-MD simulations}

Kohn-Sham DFT~\cite{Hohenberg1964,Kohn1965} is a first-principles simulation method that determines the ground state of quantum systems with high efficiency and reasonable accuracy, which has gained considerable use in computational materials science. The introduction of the Mermin scheme~\cite{Mermin1965} enabled the inclusion of excited electronic
states, which extended the applicability range of the DFT
method to higher temperatures. The combination of this method with
molecular dynamics has been widely applied to compute the EOS of
condensed matter, warm dense matter (WDM), and some dense
plasmas~\cite{Root2010,Wang2010,Mattsson2014,Zhang2018}. Unless the number of partially occupied orbitals is impractically large, DFT is typically
the most suitable computational method to derive the EOS because it accounts for electronic shell and
bonding effects. The main source of uncertainty in DFT is the use of an approximate exchange-correlation (XC) functional. The errors resulting from the XC functional often cancel between different thermodynamic conditions. Furthermore this error may only be a small fraction of the internal energy, which besides pressure is the most relevant quantity for the EOS and the derivation of the shock Hugoniot curve ~\cite{Karasiev2016}. However,
the range of validity of this assumption in the WDM regime remains to 
be verified for different classes of materials through the comparison 
with laboratory experiments and other computational technique like PIMC simulations.

With the VASP code \cite{Kresse1999}, we performed simulations from $10^4$  up to 2 million Kelvin. We employed a Nos\'e thermostat~\cite{Nose1984} to keep the temperature constant. As illustrated in Fig.~\ref{fig:EOS_ranges}, we explored densities from 6.89--51.67 \gcc~for Mg, 1--100 \gcc~for O, 2.3--18.6 \gcc~for Si, 0.35--71 \gcc~for MgO, and 0.321--64.2 \gcc~for MgSiO$_3$. We used cubic simulation cells with periodic boundary conditions and, to improve efficiency, we used a smaller number of atoms at the highest temperatures. As shown in our previous work~\cite{Driver2015b,ZhangCH2017,Driver2018,Soubiran2019}, this is not detrimental to the accuracy of the EOS data at high temperatures. More details of the simulations for BN, B, and N can be found in our previous publications~\cite{DriverNitrogen2016,ZhangBN2019,Zhang2018}.
We employed projector augmented wave (PAW) \cite{Blochl1994} pseudopotentials with a 1s$^2$ frozen core for all elements, and mostly used the Perdew-Burke-Ernzerhof (PBE) functional~\cite{PBE} at the majority of conditions to describe the exchange-correlation effects. For some materials and a small number of conditions, we had to switch to the local density approximation (see details in Ref.~\cite{Soubiran2018,Soubiran2019}), but have very good agreement between the obtained results when we switched functionals. The time step was adapted according to the corresponding temperature, and a large energy cut-off was used for the plane wave basis set.

\subsection{Shock Hugoniot Curves}
\label{hug}

The shock Hugoniot curves of many materials have been measured up to megabar, and in some cases gigabar, pressures ~\cite{Root2010,Bolis2016,Root2018,Fratanduono2018}.
Even at extreme conditions~\cite{Wang2010,Mattsson2014,ZhangBN2019,ZhangCH2018,Soubiran2019,GonzalezMilitzer2019}, predictions from first-principles simulations and experiments have been shown to be in good agreement. 

The EOS can be used to predict the thermodynamic conditions that are reached when a material is subjected to dynamical shock compression. Assuming the materials reached thermodynamic equilibrium during the experiments, the measured shock and particle velocities can be converted into pressure, density, and energy through the Rankine-Hugoniot equations~\cite{Hugoniot1887,Hugoniot1889,Ze66}. The energy conservation equation,
\begin{equation}
(E-E_0) + \frac{1}{2} (P+P_0)(V-V_0) = 0,
\label{eq:hug}
\end{equation}
is particularly convenient to derive the shock Hugoniot curve with theoretical methods.
Here, $E_0$, $V_0$, and $P_0$ represent the initial conditions of energy, volume, and pressure, respectively. 
$E$, $V$, and $P$ are the final conditions after the material behind the shock front has reached a equilibrium state. We solve the Eq.~\ref{eq:hug} for $T$ and $V$ by interpolating $E(V,T)$ and $P(V,T)$ in our EOS tables. Most simply, one solves for $V$ at given $T$ because there is only one solution.

\subsection{Linear Mixing}
The linear mixing approximation at constant pressure and temperature is the most common form in astrophysics~\cite{SC95} but it is often used also in plasma physics~\cite{Ramshaw2014}. Still, variations and alternate mixing rules have been invoked~\cite{Magyar2013}. In Ref.~\cite{Hubbard2016}, the linear mixing approximation was employed to perturb the helium fraction in an interacting H-He EOS. When mixture of carbon, oxygen, and neon nuclei are studied under conditions in White Dwarf interiors, one would want to mix the individual EOSs at constant temperature and nuclear charge density for the following reason. The density in White Dwarfs is sufficiently high for the electrons to decouple from the motion of the nuclei and to form a rigid neutralizing background. That background, however, provide the dominant contribution to the pressure and that is a function of the electron density, which is equal to the nuclear charge density. Therefore mixing at constant $P$ translates into mixing at constant nuclear charge. 

Plasmas have also been studied with two-temperatures models~\cite{Ramshaw2014} that treat nuclei and electrons as two independent thermodynamic ensembles with differing temperatures because the inter-species thermalization is delayed by the difference in mass. A number of other approaches to study mixtures~\cite{Cook2009,Cranfill2000,Clerouin2007,Horner2008,Magyar2013} have been proposed with the goal of facilitating large scale hydrodynamic simulations. Some of these approaches have been verified by orbital-free molecular dynamics~\cite{Lambert2008}.
 
In this article, however, we use the simplest form of the linear mixing approximation. For a mixture of species A and B, one neglects all inter-species interactions and, for given $P$ and $T$, one assumes the volume of the mixture is given by $V_{\rm mix}(P,T)=V_{\rm A}(P,T)+V_{\rm B}(P,T)$. The mass density, $\rho_{\rm mix}$, is given,
\begin{equation} 
{1 \over {\rho_{\rm mix}}} = {x_A \over {\rho_A}} + {x_B \over {\rho_B}},
\end{equation}
where $x_A$ and $x_B$ are the mass fractions of each species in the mixture. The internal energy is then given by 
\begin{equation} 
E_{\rm mix} = x_A E_A + x_B E_B\;,
\end{equation}
where all three energy terms are normalized per unit mass. When theoretical and computational results are employed, it is often more convenient to normalize all quantities by formula unit (FU). Let us assume that $V_1(P,T)$ and $V_2(P,T)$ are the volumes per formula unit of species 1 and 2. $N_1$ and $N_2$ specify how many formula units of species 1 and 2 are contained in one unit of the mixture. For given $P$ and $T$, the volume, mass, and internal energy of one mixture unit are obtained from,
\begin{align}
V_{\rm mix} &= N_1 V_1 + N_2 V_2\;,\label{eq:Vmix}\\
m_{\rm mix} &= N_1 m_1 + N_2 m_2\;,\\
E_{\rm mix} &= N_1 E_1 + N_2 E_2\;.
\end{align}
The mass density of the mixture is given by $\rho_{\rm mix} = m_{\rm mix} / V_{\rm mix}$.

The linear mixing approximation only provides reasonable results for the mixture at elevated temperatures where chemical bonds do not affect the EOS significantly. For this reason, we always use the $E_0$, $P_0$, and $V_0$ of the fully interacting system when we solve Hugoniot equation~\ref{eq:hug}. $E$ and $V$, however, can be approximated by $E_{\rm mix}$ and $V_{\rm mix}$. To solve Eq.~\ref{eq:hug} for a mixture, we assume a temperature and a value for $V_{\rm mix}$. Then we determine the pressure that matches $V_{\rm mix}$ and we derive the corresponding $E_{\rm mix}$. We iterate over $V_{\rm mix}$ to find a solution of Eq.~\ref{eq:hug}.

\section{Results and discussion}

\begin{figure}[htb]
    \centering
    \includegraphics[width=0.6\columnwidth]{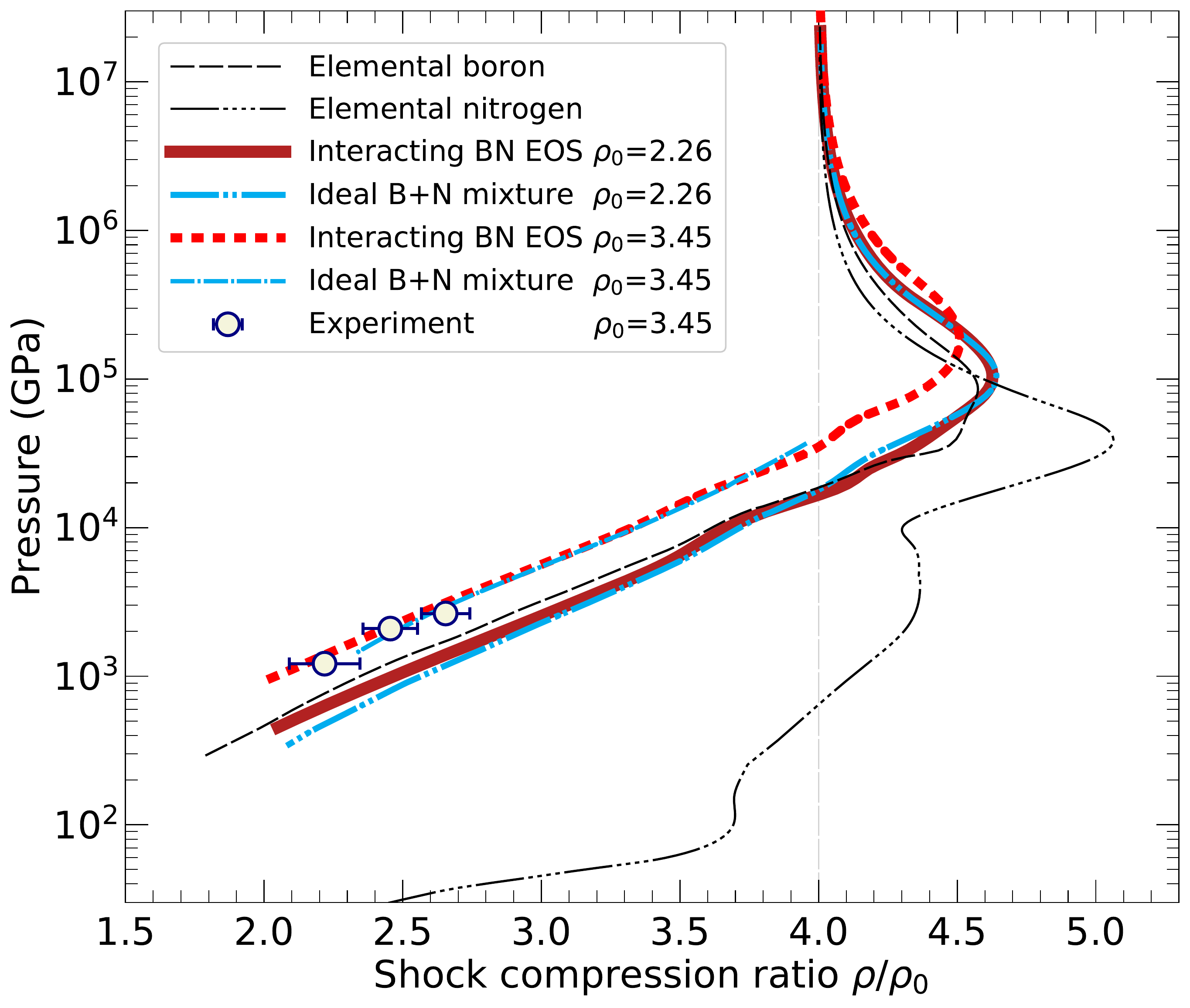}
    \includegraphics[width=0.6\columnwidth]{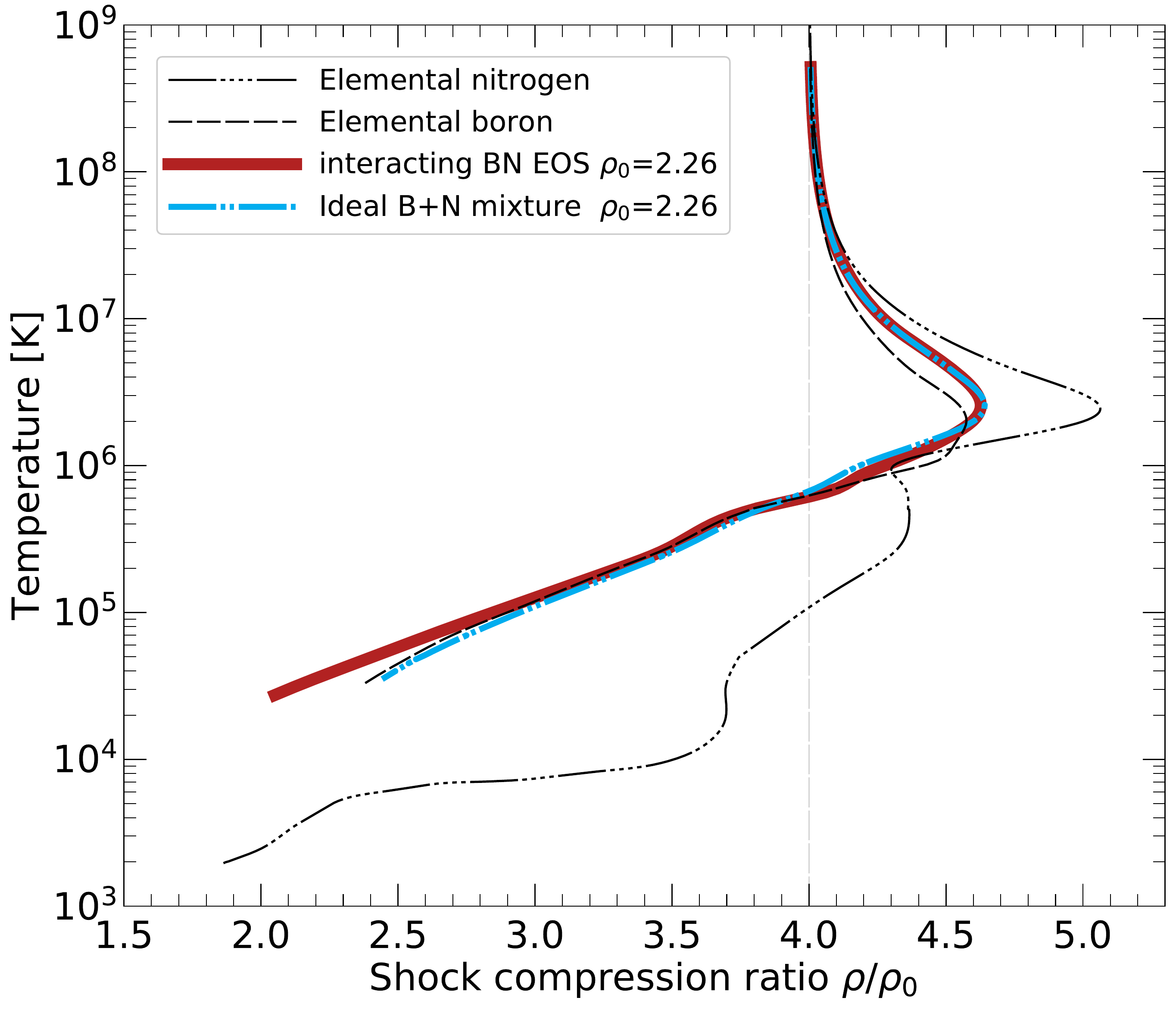}
    \caption{
    The shock Hugoniot curves of boron nitride for $\rho_0$=2.26 and 3.45 \gcc~derived from a fully interacting EOS table~\cite{ZhangBN2019} and by assuming an ideal mixture of boron~\cite{Zhang2018} and nitrogen~\cite{DriverNitrogen2016}. For temperature above $2\times 10^5\,$K, the BN Hugoniot curve is remarkably well reproduced by an ideal B+N mixture.
    Both curves are in good agreement with the experimental data from Ref.~\cite{ZhangBN2019}. For comparison, the shock Hugoniot curves of the elemental substances are also shown. In the lower panel, both bracket the Hugoniot curve of BN for the highest temperature, as is predicted by the Debye plasma model~\cite{Debye1923}. \label{fig:BN}}
\end{figure}


\begin{figure}[htb]
    \centering
    \includegraphics[width=0.6\columnwidth]{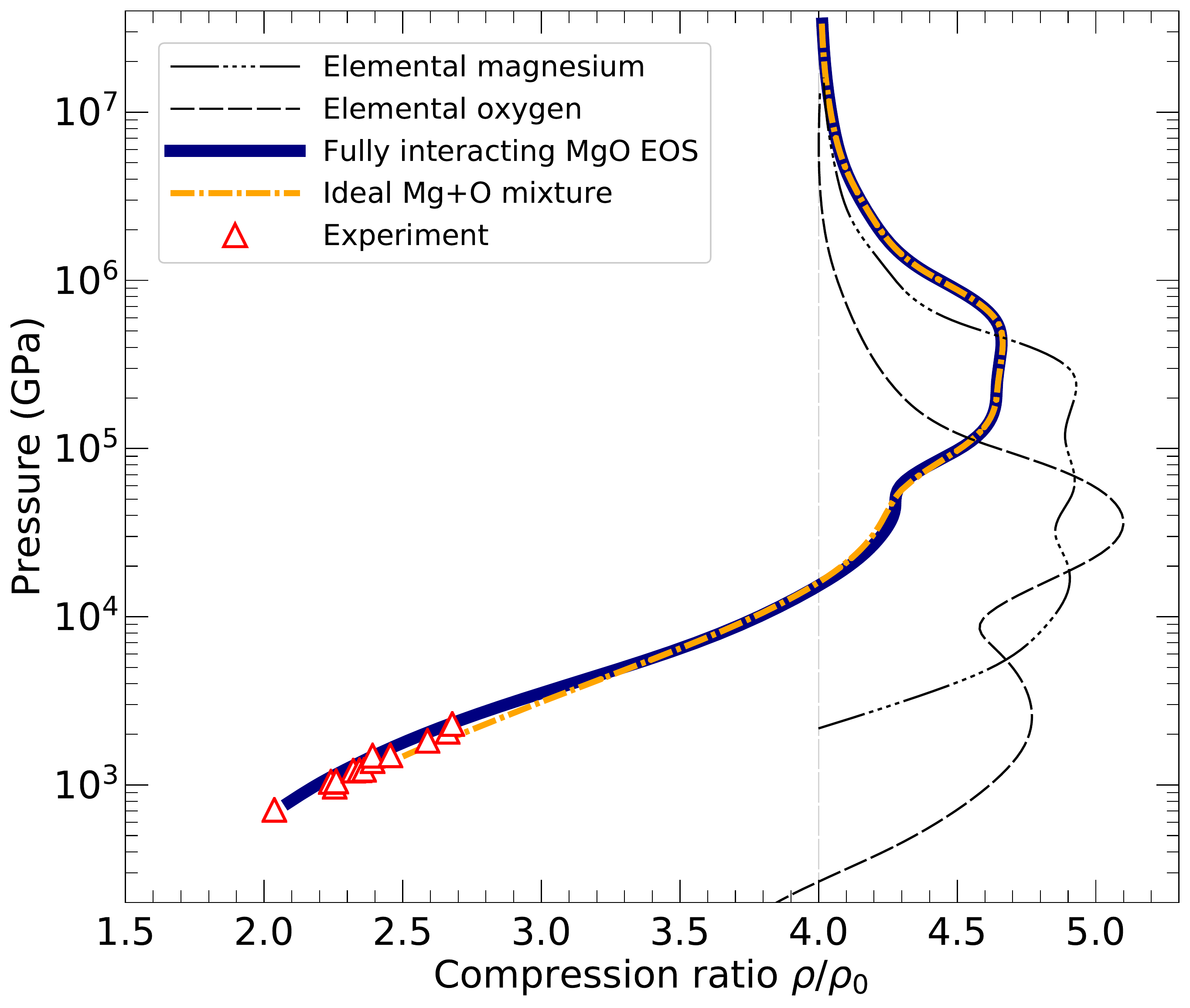}
    \includegraphics[width=0.6\columnwidth]{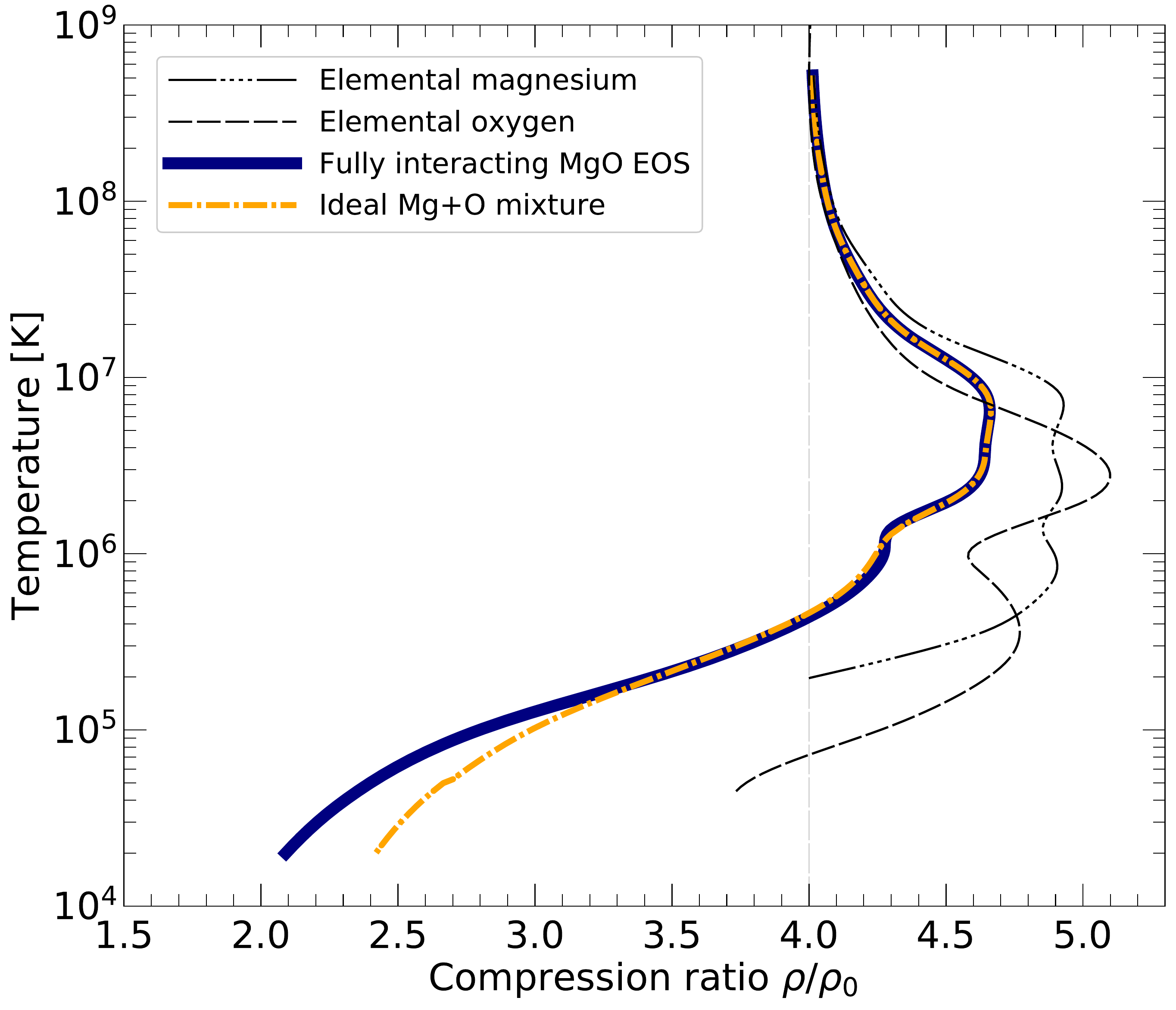}
    \caption{The shock Hugoniot curve of magnesium oxide for $\rho_0$=3.570 g/cm$^3$ derived from a fully interacting EOS table~\cite{Soubiran2019} as well as by assuming an ideal mixture of magnesium~\cite{Gonzalez-Cataldo2020} and oxygen~\cite{Driver2015b}. For comparison, the experimental results from Ref.~\cite{McCoy2019} and the shock Hugoniot curves of the elemental substances are also shown. 
\label{fig:MgO}}
\end{figure}

\begin{figure}[htb]
    \centering
    \includegraphics[width=0.6\columnwidth]{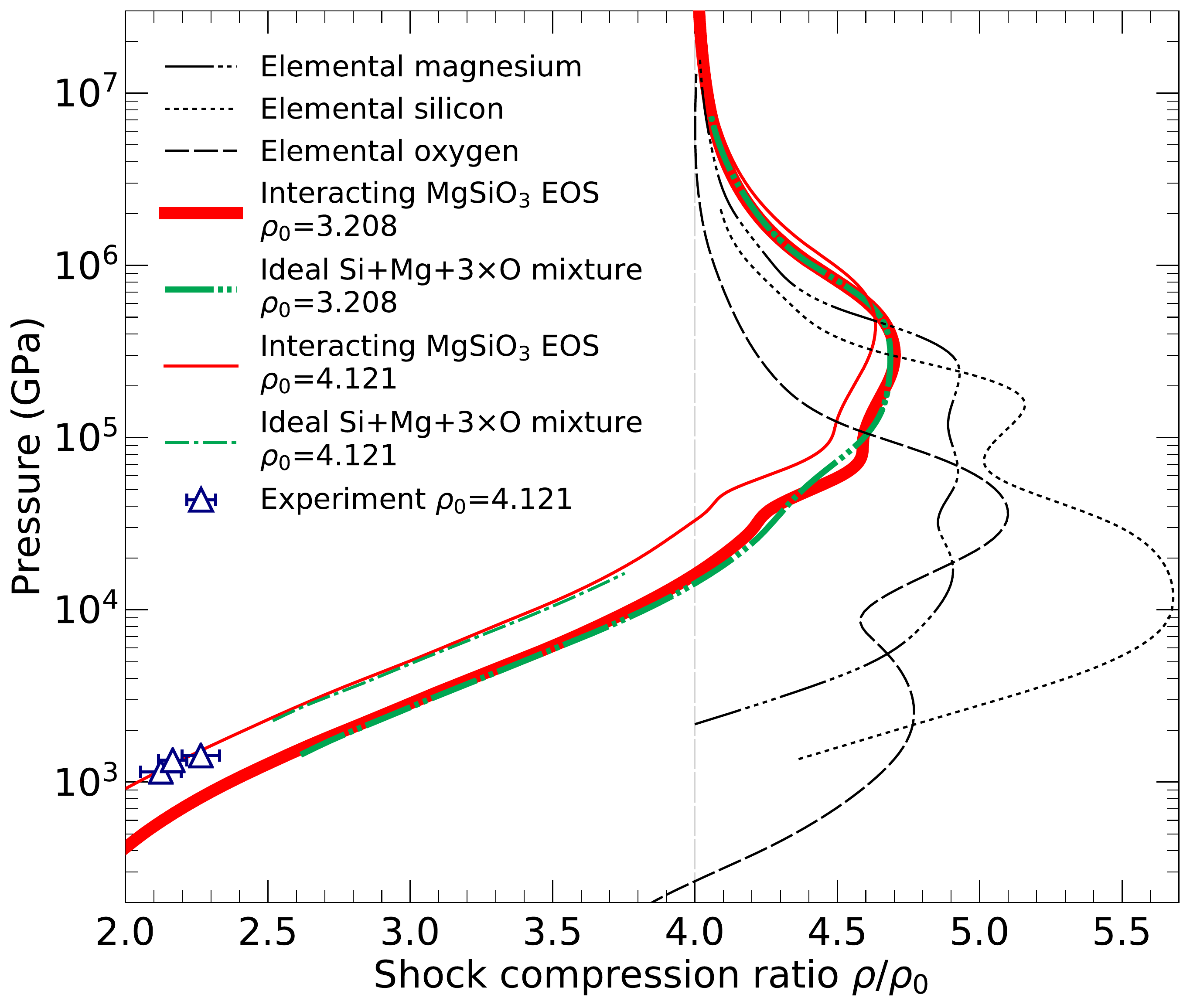}
    \includegraphics[width=0.6\columnwidth]{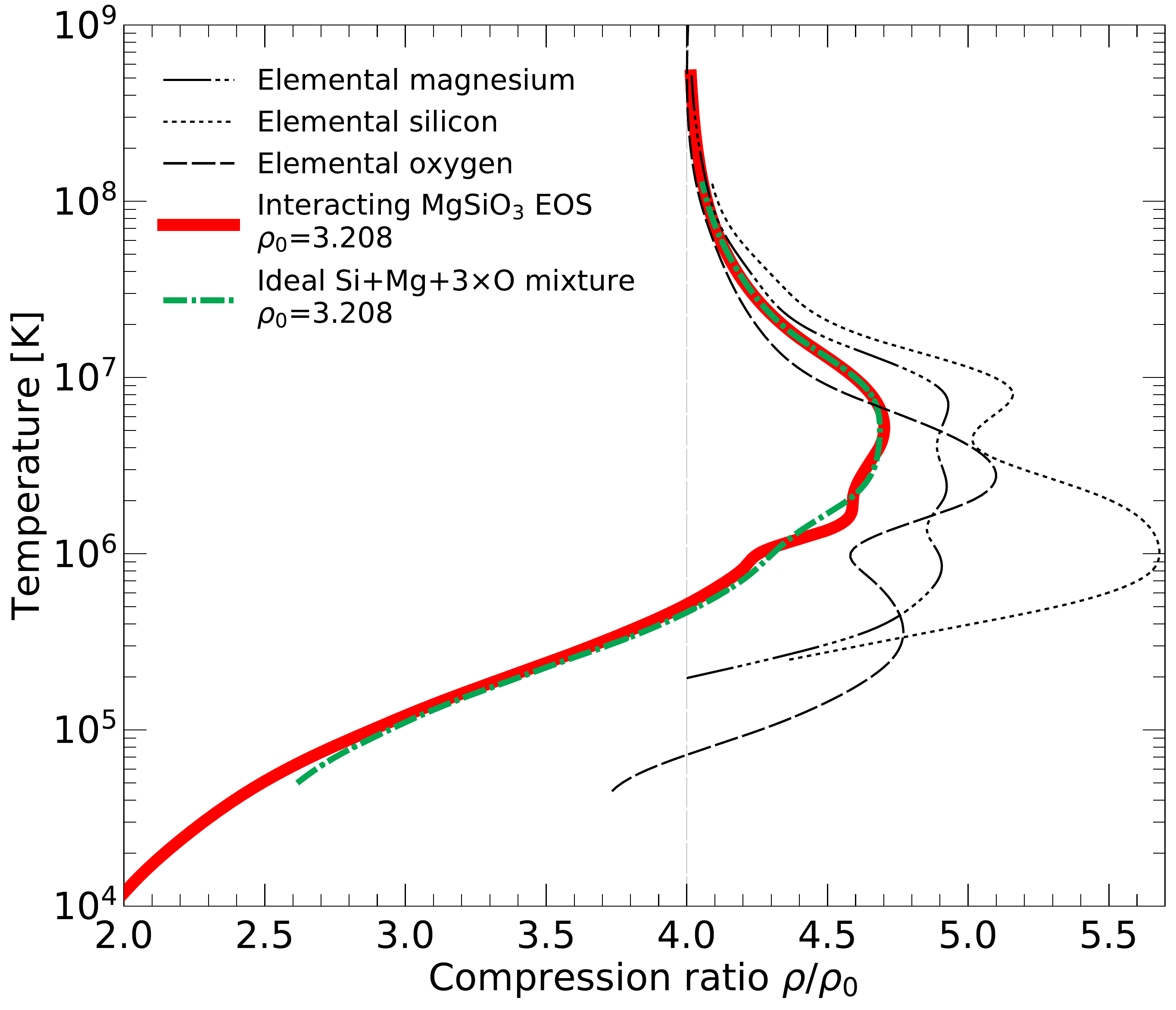}
    \caption{The shock Hugoniot curve of MgSiO$_3$ for $\rho_0$=3.208 and 4.121 \gcc~derived from a fully interacting EOS table~\cite{GonzalezMilitzer2019} as well as by assuming an ideal mixture of Mg~\cite{Gonzalez-Cataldo2020}, Si~\cite{MilitzerDriver2015}, and O~\cite{Driver2015b}. For comparison, we included the experimental data from Ref.~\cite{Millot2020} as well the shock Hugoniot curves of the three elemental substances. 
\label{fig:MgSiO3}}
\end{figure}

In Figs.~\ref{fig:BN}, \ref{fig:MgO}, and \ref{fig:MgSiO3}, we compare the shock Hugoniot curves that were computed with the fully interacting internal energy and pressure for BN~\cite{ZhangBN2019}, MgO~\cite{Soubiran2019}, and MgSiO$_3$~\cite{GonzalezMilitzer2019,GonzalezMilitzer2020} with those derived from the linear mixing approximation using the elemental EOS tables. The agreement between the pairs of curves is remarkably good in pressure-density and temperature-density spaces. For temperatures above $\sim$2$\times 10^5\,$K corresponding to shock compression ratios above $\sim$3.2, the shape of the Hugoniot curve is very well reproduced by the linear mixing approximation. This includes the regimes of K and L shell ionization. The compression maximum is also well reproduced. Linear mixtures of three elements show the same level of agreement with the fully interacting Hugoniot curves as linear mixtures of two elements. We do see some deviations for MgSiO$_3$ at 2 $\times$ 10$^6$ where the Hugoniot curves transition between the K and L shell ionization regimes. Under these conditions, the linear mixing approximation does not accurately capture the ionization equilibrium of the interacting system. 

For comparison, we also show the Hugoniot curves for the individual elements in Figs.~\ref{fig:BN}, \ref{fig:MgO}, and \ref{fig:MgSiO3}. The differences in the Hugoniots for the individual elements and that of the compounds are primarily due to differences in the initial densities. In Fig.~\ref{fig:BN}, we also show the results from laser shock experiments~\cite{ZhangBN2019} that reached up to a pressure of 2643 GPa and a compression ratio of 2.66. This is not sufficiently high for the linear mixing approximation to work well. While in pressure-compression space, the linear mixing and the fully interacting Hugoniot curves both agree the experimental data but the temperature-compression graph of Fig.~\ref{fig:BN} reveals that the shock temperatures are underestimated for compression ratios below 3.2 if the linear mixing approximation is invoked. 
We see the same trend in Fig.~\ref{fig:MgO} where we compare our theoretical predictions for MgO with the experimental results from Ref.~\cite{McCoy2019} that reached up pressures of 2303 GPa and compression ratios of 2.68. While in pressure-compression space, the predictions from the linear mixing approximations appear to be reasonable, the shock temperature is underestimated for compression ratios smaller than 3.2. Finally in Fig.~\ref{fig:MgSiO3}, we compare with the shock experiments in Ref.~\cite{Millot2020} that reached up to 1426 GPa and compression ratio of 2.26. While these results are in good agreement with fully interacting DFT-MD simulations~\cite{Millot2020}, the shock temperatures are underestimated if the linear mixing approximation is employed. The reason for this discrepancy is that, for given pressure and temperature, the linear mixing approximation underestimates the density and the internal energy for $T \lessapprox 2 \times 10^5 $K, as we confirm in the following analysis.

\begin{figure}[htb]
    \centering
    \includegraphics[width=0.6\columnwidth]{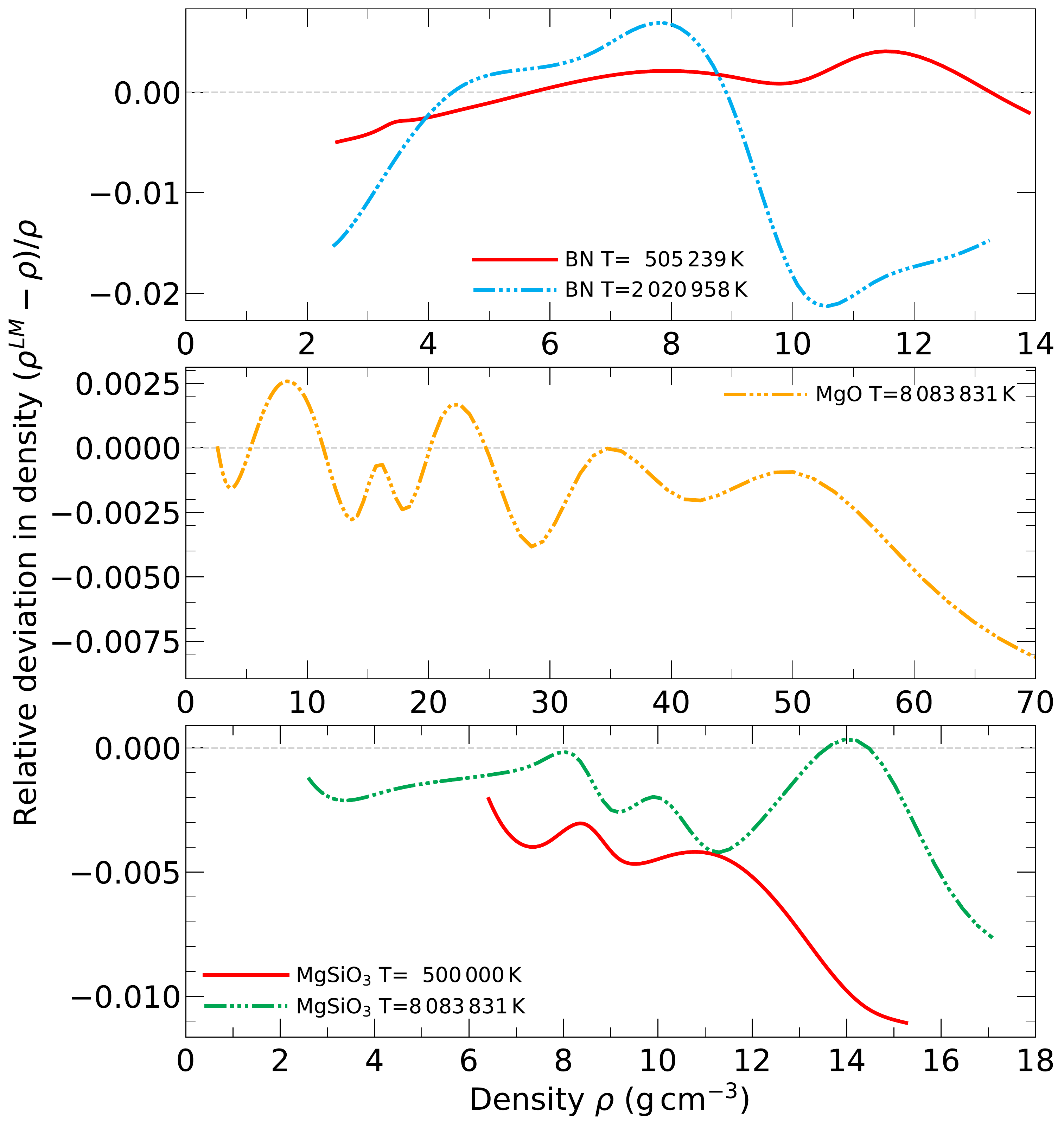}
    \caption{Relative error in the densities of BN, MgO, and MgSiO$_3$ predicted by linear mixing approximations are plotted as a function of density for two temperatures.
\label{fig:dE1}}
\end{figure}

\begin{figure}
    \centering
    \includegraphics[width=0.6\columnwidth]{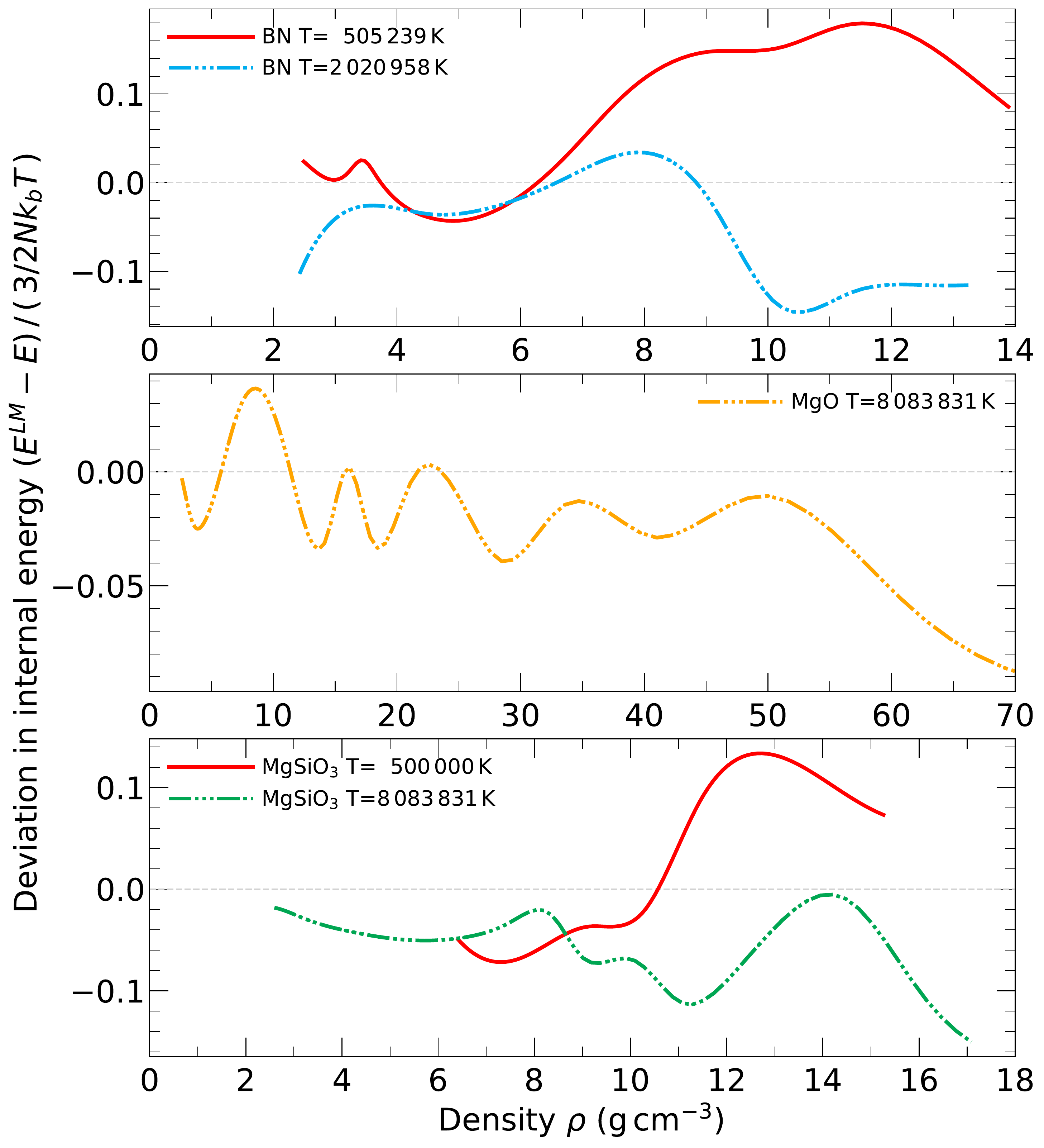}
    \caption{Error in the internal energies of BN, MgO, and MgSiO$_3$ predicted by linear mixing approximations are plotted as a function of density for the same conditions as in Fig.~\ref{fig:dE1}. 
\label{fig:dE2}}
\end{figure}

\begin{figure}[htb]
    \centering
    \includegraphics[width=0.6\columnwidth]{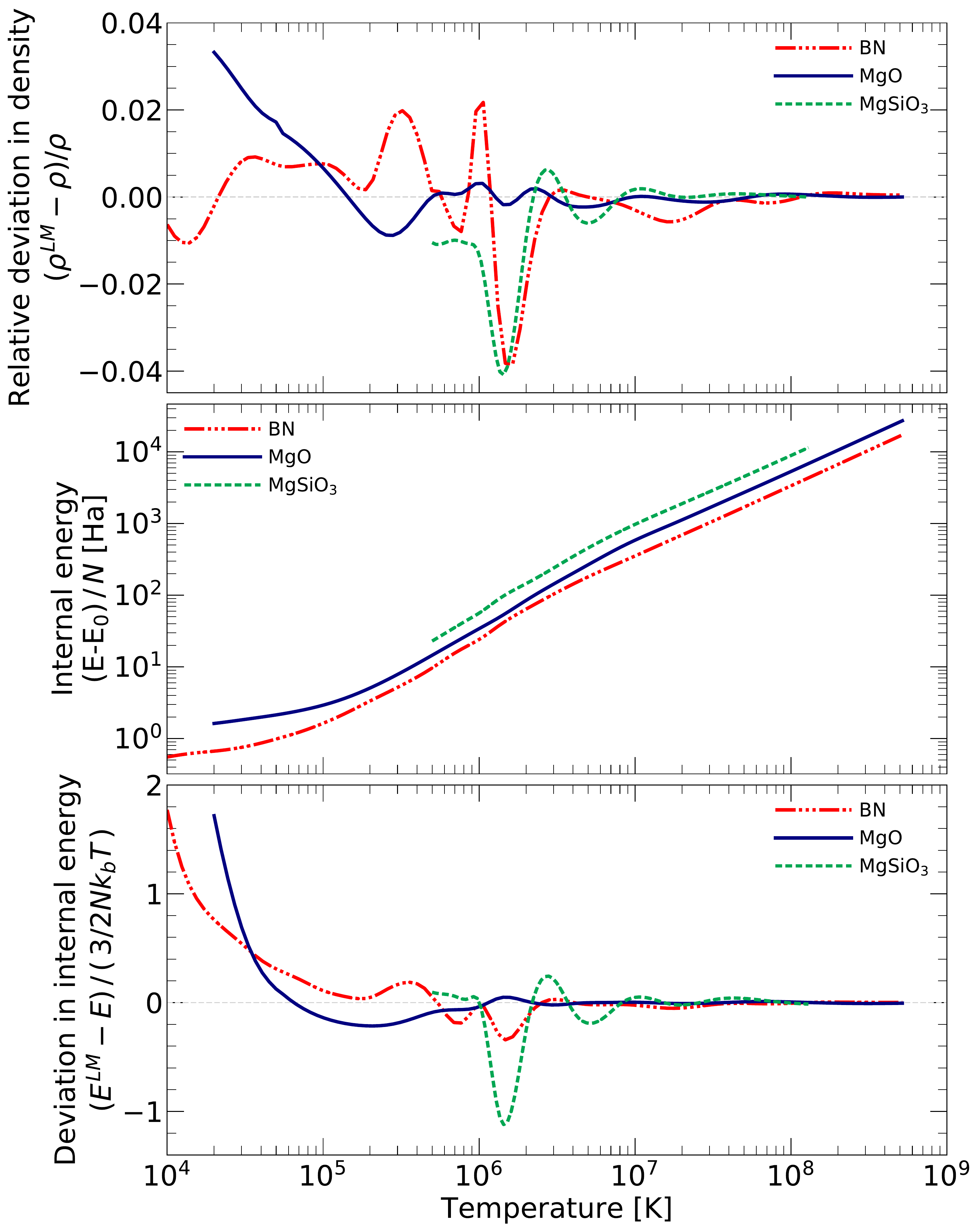}
    \caption{Errors in the linear mixing approximations are plotted as function of temperature at constant density values that were chosen to be 4.5 times the initial shock density in order to represent conditions near the maximum shock compression. For BN, MgO, and MgSiO$_3$ the densities were respectively 10.161, 16.064, and 14.436 \gcc. The top panel shows relative error in the predicted density. In the middle panel, we show the internal energy per atom on a logarithmic scale after subtracting $E_0$. In the lower panel, we show 
    energy error in the linear mixing approximation that we normalized by the ideal kinetic energy, $\frac{3}{2} N k_\text{B} T$. }
    \label{fig:dE3} 
\end{figure}

The shock Hugoniot curves can only be reproduced well by the linear mixing approximation as long as the volume and internal energy, that enter into Eq.~\ref{eq:hug}, are reasonably accurate. In Fig.~\ref{fig:dE1}, we plot the deviation in $\rho_{\rm mix}$ and the interacting $\rho$ for two temperatures as a function of density. The error in density is less than 1\% for all three materials with the exception of a B+N mixture, in which case the error reaches 2\% at low and high densities. 

In Fig.~\ref{fig:dE2}, we plot the linear mixing error in the internal energy that we normalized by dividing by nuclear kinetic energy $\frac{3}{2} N k_\text{B} T$. The deviations are 0.1 or less for all three materials and conditions under consideration. 

In Fig.~\ref{fig:dE3}, we plot the linear mixing errors as a function of temperature for three relevant density set equal to 4.5 times the initial shock density, $\rho_0$. The deviations in density are 2\% or less except at 1.5 $\times 10^6$ K where we switch between PIMC and DFT-MD EOS computations, in which case the deviations reaches 4\%. This, however, does not reflect any insufficiency in the linear mixing approximation but the underlying EOS tables are imperfect. When we study the linear mixing error in the internal energy, we also find a discrepancy for MgSiO$_3$ at 1.5 $\times 10^6$ K. 

For temperature below $2 \times 10^5$ K, the errors in density and internal energy of the linear mixing approximation increase with decreasing temperature because chemical bonds and interactions between different species play an increasingly important role. Chemical bonds lower the internal energy and pressure. Since bonding effects are absent from the linear mixing approximation, it overestimates the internal energy and density for given pressure and temperature, which explains the trends at lower temperature in Fig.~\ref{fig:dE3}. Still, already for T $\gtrapprox 2 \times 10^5$ K, the linear mixing approximation works well.

\begin{table}[h]
  \caption{For the materials and initial conditions in the first two columns, the three following column pairs tabulate what changes in pressure or internal energy are needed to a) shift the maximum compression ratio on the Hugoniot curve by $-$0.1 or shift the $\rho/\rho_0=3.5$ point b) by 5\% up in temperature and c) by 5\% up in pressure.}
  \label{tab1}
\begin{center}
\begin{tabular}{ c| c| l l|l l|l l } 
  Material & $\rho_0$ (\gcc) 
  & $\frac{\delta P_{\rm max}}{P}$ & $\frac{\delta E_{\rm max}}{\frac{3}{2} N k_b T}$
  & $\frac{\delta P_{1}}{P}$ & $\frac{\delta E_{1}}{\frac{3}{2} N k_b T}$
  & $\frac{\delta P_{2}}{P}$ & $\frac{\delta E_{2}}{\frac{3}{2} N k_b T}$
  \\
 \hline
 BN  & 2.26  &  0.030 & $-$0.206 &  0.011 & $-$0.039 &  0.009 & $-$0.041 \\
 MgO & 3.570 &  0.029 & $-$0.336 &  0.018 & $-$0.085 &  0.015 & $-$0.097 \\
 MgSiO$_3$ &  3.208 &  0.029 & $-$0.324 &  0.019 & $-$0.085 &  0.015 & $-$0.095 \\
 \end{tabular}
\end{center}
\end{table}

Finally, we performed three tests how much of a change in pressure and in internal energy is needed to shift the Hugoniot curve in temperature-compression and pressure-compression spaces. First, we determined what fractional change in pressure is needed to reduce the maximum compression ratio on the three principal Hugoniot curves by 10\%. The results of our calculations in Tab.~\ref{tab1} show that a 3\% increase in pressure would trigger such a shift. Alternatively, such a reduction in compression ratio can be introduced by lowering the internal energy by 0.21$\ldots$0.34 $\times \frac{3}{2} N k_b T$. The magnitude of both corrections is much larger than the linear mixing error that we reported in Figs.~\ref{fig:dE1},~\ref{fig:dE2}, and~\ref{fig:dE3}, which explains why we were able to reproduce the compression maxima of the Hugoniot curves very well with the linear mixing approximation. 

In Tab.~\ref{tab1}, we report the results from two more tests at lower temperatures and pressures. Starting the point at 3.5-fold compression on the Hugoniot curve, we asked what fractional pressure change and what energy correction in units of $\frac{3}{2} N k_b T$ would be needed to move $\rho/\rho_0=3.5$ point up in temperature by 5\% or up in pressure by 5\%. The required pressure and energy corrections are reported in columns 5-8 of Tab.~\ref{tab1}. Energy changes between $-0.097$ to $-0.039 \times \frac{3}{2} N k_b T$ are needed to change the temperature and pressure on the Hugoniot curve by 5\%. These changes are comparable in magnitude to the linear mixing errors we have reported in Figs.~\ref{fig:dE2} and \ref{fig:dE3}. So, at 3.5-fold compression, the accuracy of the linear mixing approximation is about 5\%. 

In Tab.~\ref{tab1}, we also show that fraction pressure changes between $0.009$ and $0.019$ are needed to move $\rho/\rho_0=3.5$ point by 5\% in pressure and temperature. These values are larger than the linear mixing errors in density at 5$\times 10^5$ K that we show in Fig.~\ref{fig:dE1} that amount to less than 1\% at this temperature. This suggest that the changes in the internal energy are slightly more difficult to reproduce with the linear mixing approximation than the pressure.

\section{Conclusion}

We have validated the linear mixing approximation across a wide range of temperature and pressure conditions for MgO, MgSiO$_3$, and BN plasmas. Under this approximation, accurate shock Hugoniot curves can be obtained for temperatures of $T \gtrapprox 2 \times 10^5$ K and compression ratios of $\rho/\rho_0 \gtrapprox 3.2$, correctly predicting the maximal compression ratio and the K- and L-shell ionization regimes. This will greatly simplify the computations for the regime of WDM and may even help reduce the number of experiments. This conclusion is further supported by the first-principles calculations for CH~\cite{ZhangCH2017} that reported that the maximal compression ratio on the Hugoniot curve can be derived with an accuracy of 1\% being combining the EOSs of elemental carbon and hydrogen. Similarly, we are able to reproduce the maximum compression ratio of B$_4$C~\cite{Zhang2020} with an accuracy of 0.4\% if we mix the EOS of boron and carbon. Ref.~\cite{Soubiran2015} determine the hydrogen and water form an ideal mixture under conditions of ice giant envelopes. On the other hand, mixtures of hydrogen, helium and heavier elements in giant planet envelopes could only be accurately represented as a linear mixture after the volumes of the heavier species had been adjusted to match the results from fully interacting DFT-MD simulations~\cite{Soubiran2016}. The validity of the linear mixing approximation depends on the pressure and temperature conditions. Magyar and Mattsson showed that errors of 10\% can be expected for xenon-deuterium mixtures at megabar pressures and 10$\,$000 K~\cite{Magyar2013} while we have shown here that at much higher temperature, the linear mixing approximation works very well. Still, in this article we only investigate how well the volume of and internal energies can be derive by combining the EOSs of the various elements at constant pressure and temperature. Future work should thus be directed on understanding to which degree transport properties of WDM are affected by nonlinear mixing effects.

\begin{acknowledgments}
  This work was in part supported by the National Science
  Foundation-Department of Energy (DOE) partnership for plasma science and engineering (grant DE-SC0016248) and by the DOE-National Nuclear Security Administration (grant DE-NA0003842). Part of this work was performed under the auspices of the U.S. DOE by Lawrence Livermore National Laboratory under Contract No. DE-AC52-07NA27344. Computational support was provided by the Blue Waters sustained-petascale computing project (NSF ACI 1640776) and the National Energy Research Scientific Computing Center (NERSC).
  
  HDW, DCS and, MM state that their work was sponsored by an agency of the United States government. Neither the United States government nor any agency thereof, nor any of their employees,
makes any warranty, express or implied, or assumes any legal
liability or responsibility for the accuracy, completeness, or
usefulness of any information, apparatus, product, or process
disclosed, or represents that its use would not infringe privately owned rights. Reference herein to any specific commercial product, process, or service by trade name, trademark,
manufacturer, or otherwise does not necessarily constitute
or imply its endorsement, recommendation, or favoring by
the U.S. Government or any agency thereof. The views and
opinions of authors expressed herein do not necessarily state
or reflect those of the U.S. Government or any agency thereof, and shall not be used for advertising or product endorsement purposes.  
\end{acknowledgments}

\section*{Data availability}
The data sets of all figures are available from the corresponding author upon request.



\providecommand{\noopsort}[1]{}\providecommand{\singleletter}[1]{#1}%

\end{document}